# Incentivising Privacy in Cryptocurrencies

Sarah Azouvi[*]
University College London
sarah.azouvi.13@ucl.ac.uk Alexander Hicks[*]
University College London
alexander.hicks@ucl.ac.uk Haaroon Yousaf[*]
University College London
h.yousaf@ucl.ac.uk

January 9, 2019

This work was presented as a poster at OPERANDI 2018. The presented poster is the final page of this document.

## 1 Introduction

Privacy was one of the key points mentioned in Nakamoto's Bitcoin whitepaper [9], and one of the selling points of Bitcoin in its early stages. In hindsight, however, de-anonymising Bitcoin users turned out to be more feasible than expected [7]. Since then, privacy focused cryptocurrencies such as Zcash and Monero have surfaced. Both of these examples cannot be described as fully successful in their aims, as recent research has shown [6, 8, 10].

Incentives are integral to the security of cryptocurrencies [2], so it is interesting to investigate whether they could also be aligned with privacy goals. A lack of privacy often results from low user counts, resulting in low anonymity sets. Could users be incentivised to use the privacy preserving implementations of the systems they use? Not only is Zcash much less used than Bitcoin (which it forked from), but most Zcash transactions are simply transparent transactions, rather than the (at least intended to be) privacy-preserving shielded transactions. Our aim is to discuss how incentives could be incorporated into systems like cryptocurrencies with the aim of achieving privacy goals. We take Zcash as example, but the ideas discussed could apply to other privacy-focused cryptocurrencies.

## 2 Background

### 2.1 Privacy in Zcash

Zerocash [11] was proposed as a privacy focused alternative to Bitcoin, based on recent advances in zero-knowledge proof techniques. It was implemented as a fork of Bitcoin, called Zcash. Marketed as "the first open, permissionless cryptocurrency that can fully protect the privacy of transactions using zero-knowledge cryptography", it has recorded over two million transactions, and reached a billion dollar market cap. To address Bitcoin's privacy problem, Zcash adds a shielded pool that allows users to obscure parts of a transaction. To interact with the pool, Zcash includes three new types of transactions: shielded transactions that hide the recipient, private transactions that hide the sender, recipient and value, and de-shielded transactions that hide the sender. Standard, fully public transactions like those in Bitcoin are also possible. Table 1 summarises the hidden and revealed parts of transactions.

Unfortunately, privacy guarantees turn out to be weaker than might be desired. Analysis of over two million transactions reveals that the vast majority of transactions, 83.8%, were simply standard public transactions (Figure 1), reducing the size of the anonymity set. Applying simple heuristics further reduces the anonymity of users [6, 10].

---

[*]These authors contributed equally to this work.

Table 1: Attributes revealed in the different types of Zcash transactions.

|              | Sender | Receiver | Amount |
|--------------|:------:|:--------:|:------:|
| Public       | ✓      | ✓        | ✓      |
| Shielded     | ✓      | ✗        | ✓      |
| Private      | ✗      | ✗        | ✗      |
| De-shielded  | ✗      | ✓        | ✓      |

## 2.2 Incentives and security of cryptocurrencies

Cryptocurrencies are examples of systems that inherently rely on incentives, like mining rewards, that provide part of the network's security. This has had clear success, to the extent that no major cryptocurrency has been irreversibly impacted by dishonest miners. Nonetheless, many of the known issues in the security of cryptocurrencies are related to incentives [3–5], showing that despite existing success their use is not yet fully understood.

## 3 Discussion

The aim of the talk will be to discuss different ways of increasing the privacy guarantees Zcash provides, focusing on the role incentives play in deploying these.

The problem can be understood in terms of the three parties involved: users, the network (i.e., the consensus protocol and participating miners) and the governance (i.e., influential members of the Zcash company, affiliates and important miners). For users, the goal is to perform transactions that are quickly verified, cheap and private. For miners participating in the consensus protocol, the goal is to maximize their short term-profit whilst ensuring long term health of the network (the so called *tragedy of the commons* [3]). The governance aims to ensure long term health of the network, and uses its influence to enforce changes.

The first problem is that users are not actually performing private transactions. How could they be incentivised to do so? One idea is simply to offer some form of reward for the use of private transactions. Users as a whole may be ready to subsidise this type of reward. Frequent users, who would prosper more from the increase in privacy, would pay out the most to the rest of the active network, while less frequent users would pay less but be encouraged to use private transactions nonetheless through the reward. This change would be hard to implement and places the cost of the privacy on the user side. How much would users be willing to pay for a collective improvement in privacy?

Another solution would be to incorporate the incentivisation within the protocol rather than having users pay for it. Mining rewards already exist, so it could simply be slightly modified to offer more to miners that include private transactions in their block. Miners would then prioritise these, leading users to favour these so that their transactions are more likely to quickly be included in blocks. As a result users would also be "rewarded" by doing private transactions as this means they could pay less transaction fees when using the privacy feature. Indeed as miners are already incentivized to prioritize these transactions, they would include them in their blocks even if their transaction fees are lower than non-shielded transactions. In addition this would mean that non-shielded transactions would probably need to pay higher transaction fees to get included, which again disincentivises a user to use those. We could also add an explicit extra-fee for non-shielded transactions to enforce the same behaviour but that may be a bit of an overkill. Adding such a reward would only require a small change to the protocol, at no cost to the users, so it might appear like an easy choice for the governance to make. There are some drawbacks.

For one, the reward for private transactions would either be added on top of the existing reward, increasing inflation, or be part of the existing reward. In this case, the mining reward would be split between the block reward and a privacy reward, that could increase with each private transaction added, up to the full mining reward. But miners may then be unhappy that their mining rewards could end up being lower. There also exists a founders reward, that has so far sent 20% of all created coins to founders, although the number of coins given to founders should reach a cap in the next years. Redirecting part of these funds to privacy rewards would then be another option.

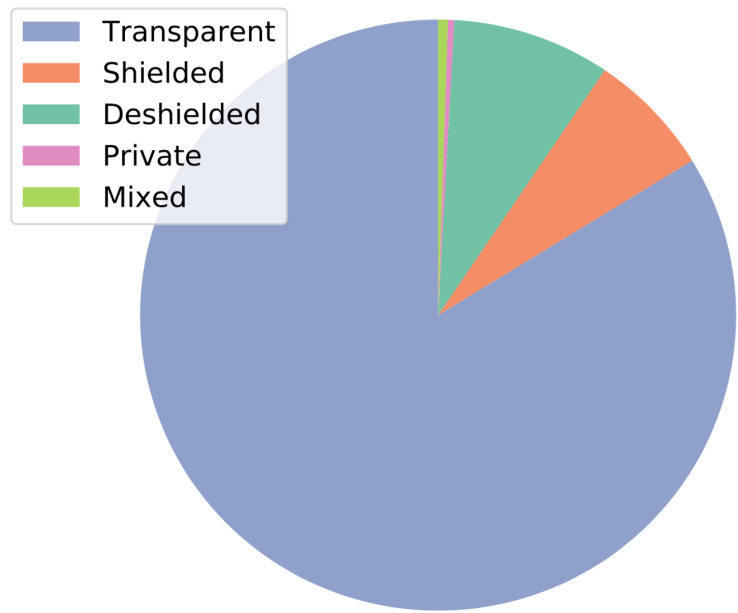

Figure 1: The percentage of types of transactions in Zcash from the start of the chain to April 2018.

There are also computational and usability aspects to take into account. Private transactions are of a higher size, and take much more computational power to produce. At present this is costly to do, especially on commodity hardware. There is, however, an upcoming update (Sapling) that should improve the speed and reduce the computational cost of private transactions.

Enforcing use of the shielded pool could also be done, leaving only private transactions. This, however, raises some issues. How could the governance impose an update to move all coins the pool? Should all public coins be locked until they are used in the pool? There are also issues of fungibility and ownership. It could make it harder for exchanges to migrate to a private coin, as its nature may make trading and auditing very difficult. Full privacy raises the difficulty for law enforcement to trace coins used for illegal purposes, which could lead to regulatory issues. Anderson et al. have also argued that coins received from those mixes could have been previously tainted, raising legal and ownership issues [1].

Finally, there are also usability issues that could be addressed. Low adoption of private transactions can be explained by low levels of support for these in wallets, particularly mobile ones, and exchanges. As far as we know, WinZec is the only wallet service that supports use of the pool, but it is only available on Microsoft Windows. Users must otherwise use a command line interface, that is clearly not ideal for users that may have no experience doing so. Could there be incentives for the Zcash company, or the community, to work towards better usability in order to further their privacy goals?

# Incentivising Privacy in Cryptocurrencies

Sarah Azouvi, Haaroon Yousaf and Alexander Hicks

OPERANDI 2018: Open Day for Privacy, Transparency and Decentralization

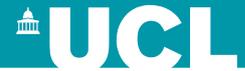

## What is wanted?

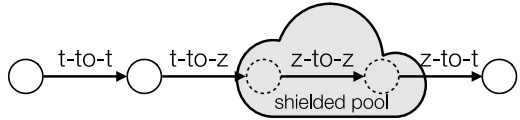

A simple diagrams illustrating the different types of Zcash transactions

- Users
  - Perform fast, quickly verified, cheap and private transactions
- Miners
  - Maximise short term profit and long term health of the network (*tragedy of the commons* [4])
  - Regular flow of transactions to benefit from fees
- Governance
  - Long term network health
  - Enforce changes to design, protocol and additional features

|             | Sender | Receiver | Amount |
|-------------|--------|----------|--------|
| Public      | ✓      | ✓        | ✓      |
| Shielded    | ✓      | ✗        | ✓      |
| Private     | ✗      | ✗        | ✗      |
| De-shielded | ✗      | ✓        | ✓      |

Data revealed per transaction type

## Current problems with Zcash

- Few private transactions (16.2%) [1]
- Low support for private transactions on clients
- Generating private transactions requires heavy computation power, limited on mobile devices
- Longer verification than public, increasing wait times
- Reduced use results in smaller anonymity pool
- Combined with metadata analysis allows de-anonymisation [1-3]
- Limited incentives to use costly private transactions
- Various types of transactions create confusion
- Standard practise given but not enforced

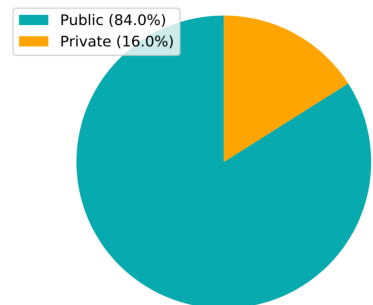

Percentage of types of transactions

## Potential solutions to incentivise private transactions

- Increase transaction fees for non-private transactions

- Reduce transaction fees for private transactions

- Increase mining reward proportionally to number of private transactions within new blocks
  - May cause an increase in inflation or reduction in existing rewards

- Enforce use of the private pool
  - Difficult to move all coins into the pool
  - Potential lock-down of public coins, unlocked when moved into the pool
  - Reduced fungibility and ownership
    - Difficult for exchanges, law enforcement and services to identify tainted coins
    - Full privacy may cause regulatory issues [5]

- Improve usability
  - Official cross-platform client from the developers
  - Simpler command line interface

| Type        | Average Fee |
|-------------|-------------|
| Public      | 0.0051      |
| Shielded    | 0.0009      |
| Private     | 0.0001      |
| De-shielded | 0.0001      |

Average transaction fee (ZEC) per type

Supported in part by EPSRC Grant EP/N028104/1, in part by the EU H2020 TITANIUM project under grant agreement number 740558 and by X Y Z